# Shock dynamics of strong imploding cylindrical and spherical shock waves with non-ideal gas effects

**R. K. Anand**

**Department of Physics, University of Allahabad, Allahabad-211002, India**

**Abstract** In this paper, the generalized analytical solution for one dimensional adiabatic flow behind the strong imploding shock waves propagating in a non-ideal gas is obtained by using Whitham's geometrical shock dynamics theory. Landau and Lifshitz's equation of state for non-ideal gas and Anand's generalized shock jump relations are taken into consideration to explore the effects due to an increase in (i) the propagation distance from the centre of convergence, (ii) the non-idealness parameter and, (iii) the adiabatic index, on the shock velocity, pressure, density, particle velocity, sound speed, adiabatic compressibility and the change in entropy across the shock front. The findings provided a clear picture of whether and how the non-idealness parameter and the adiabatic index affect the flow field behind the strong imploding shock front.

**1. Introduction**

The study of shock waves is an important research field for the safety assessments and predictions of disasters due to explosions. Shock waves are produced due to the sudden release of enormous amount of energy from sources such as a nuclear explosion, a chemical detonation, or rupture of a pressurized vessel. These waves are characterized by a supersonic shock front followed by an exponential-type decay of the physical properties of the gas [1]. Shock waves are common in the interstellar medium because of a great variety of supersonic motions, supernova explosions; central part of star burst galaxies, etc. Imploding shock waves have been a field of continuing research interest over the years as possible methods for generating high-pressure, high-temperature plasmas at the centre of convergence, as well as to understand the basic fluid dynamics involved in the process. Since at the centre of convergence all physical variables tend to infinity, numerical methods fail to make predictions about the occurrences in the immediate vicinity of the centre. Shock waves are usually generated by point explosions (nuclear explosions and detonation of solid explosives, solid and liquid propellants rocket motors), high pressure gas containers (chemical explosions) and laser beam focusing.

Shock wave phenomena also arise in astrophysics, hypersonic aerodynamics and hypervelocity impact. An understanding of the properties of the shock waves both in the near-field and the far-field is useful with regard to the characteristics such as shock strength, shock overpressure, shock speed, and impulse.

The first study on converging shock waves was performed by Guderley [2], who presented his well-known self-similarity solution of strong converging cylindrical and spherical shock waves close to focus. The problem was also studied independently by Chester [3], Chisnell [4], and Whitham [5] with approximate methods, specifically geometrical shock dynamics. In their solutions, the exponent in the expression for Mach number as a function of shock radius for the spherical case is exactly twice that for the cylindrical case. This approximate result differs from the exact solution by less than one percent. The geometrical shock dynamics approach is both simple and intuitive while providing fairly accurate results. Sakurai [6, 7], Sedov [8], Zel'dovich and Raizer [9], Lazarus and Richtmeyer [10], Van Dyke and Guttmann [11] and Hafner [12] presented high-accuracy results adopting alternative approaches for the investigation of the implosion problem under consideration. A numerical solution for a converging cylindrical shock wave was presented by Payne [13]. The problem of contracting spherical or cylindrical shock front propagation into a uniform gas at rest was studied by Stanyukovich [14]. The effects of overtaking disturbances on the motion of converging shock waves were studied by Yousaf [15]. The problems of implosion of a spherical shock wave in a gas and collapse of a spherical bubble in a liquid are discussed by Zel'dovich and Raizer [9] by using a self similar solution method. Matsuo [16] has described the history of the fluid from the initial stage to the focusing stage by using non-similar approximate method. Chisnell [17] gave an analytical description of the flow behind converging shock waves through a study of the singular points of the differential equations and investigated the shock behavior when the specific heat ratio, tends to zero or to infinity. Ponchaut *et al* [18] obtained universal solutions for initially infinitesimally weak imploding cylindrical and spherical shock waves in a perfect gas. Gurovich, Grinenko and Krasik [19] obtained the semi-analytical solution for the problem of converging shock waves. A comparison of the solutions of self-similar theory, geometric shock dynamics, as well that of a numerical Euler solver was presented by Hornung, Pullin and Ponchaut [20], showing good agreement. The Interaction of a Taylor blast wave with isotropic turbulence was analyzed by Bhagatwalaa and Lele [21]. Recently, Bhagatwalaa and Lele [22] studied the interaction of converging spherical shock waves with isotropic turbulence.

Due to the extremely high temperatures, particularly at the beginning of shock propagation, the assumption of a perfect gas is not valid. As the strength of the converging shock increases the non-ideal gas effects become significant and must be included in the theoretical investigations. These effects need to be accounted for in order to correctly describe the post shock conditions and acquire information on the attainable pressure and temperatures by shock focusing. The study of imploding shock waves in real gases is of immense significant due to its wide applications to supersonic flights in polluted air, metalized rocket propellants, bomb blasts, coal mine blasts, lunar ash flow, nozzle flow, underground, volcanic and cosmic explosions, collision of comet with a planet and many other astrophysical situations and engineering problems in industry and the environment. The motion of spherical and cylindrical shock waves in real gases were studied by Steiner and Gretler [23] taking into account the high temperature effects of vibration, dissociation, electronic excitation, and ionization, as well as the intermolecular forces at high pressures. The flow field behind the converging shock waves was investigated in van der Waals gas by Wu and Roberts [24, 25] and Evans [26]. Using the direct simulation Monte Carlo (DSMC) method, Goldsworthy and Pullin [27] investigated the effects of finite Knudsen number in the problem of a cylindrically imploding shock wave and found that the solution of the dissipative flow field displays a scaling consistent with the Guderlay similarity exponent. Kjellander, Tillmark and Apazidis [28] studied strong cylindrical and spherical shock implosion in a monatomic real gas and shown that ionization has a major effect on temperature and density behind the converging shock as well as on the shock acceleration. Recently, Zhigang *et al* [29, 30] proposed a simple but effective technique based on shock dynamics to generate cylindrical converging shock waves. The theoretical investigations adopting above mentioned approaches are in close agreement with the experimental results [31-33]. Thus, the dynamics of the propagation of imploding shock waves is accurately predicted by the power-law solution, and geometrical shock dynamics approach [34].

To author's best knowledge, so far there is no paper reporting the analytical solutions for imploding shock waves propagating in a non-ideal gas taking into consideration the equation of state as given by Landau and Lifshitz [35]. The geometrical shock dynamics approach is employed to investigate the motion of converging shock waves and it gives highly accurate results especially, in the case of a spherical symmetry. Since a converging shock is strengthened as it focuses on the origin, the strong-shock approximation is appropriate. This has developed my interest in studying the strong imploding shock waves propagating in non-ideal gas. Such a problem is of great interest in astrophysics as it is highly relevant to the problem of the origin of cosmic rays [6, 7, 9, 12].

Using a numeric method of characteristics Steiner and Gretler [23] found substantial deviation in the initial stages of the explosion from the ideal gas solution due to the inclusion of conductive and radiative heat transfer, in particular the variations in temperature profiles and the thermodynamic state just behind the imploding front. Kjellander, Tillmark and Apazidis [28] found considerable decrease in temperature and increase in density as compared to the ideal nonionizing case; whereas the ionization does not have the same striking effect on pressure behind the imploding front. In the present basic research paper, the author has investigated the effects of non-idealness parameter and adiabatic index of the gas on the flow variables just behind the strong imploding shock front. For this purpose, a geometrical shock dynamics model is developed to provide a simplified, complete treatment of the propagation of strong imploding cylindrical and spherical shock waves in non-ideal gases during the convergent process. Geometrical shock dynamics was introduced by Whitham [5] and the method is further described in Whitham [34]. The original geometrical shock dynamics method does not account for the influence of the flow ahead of the shock. The equation of state used is Landau and Lifshitz's [35] model for a non-ideal gas. It is worth mentioning that the effects due to the non-ideal gas enter through the non-idealness parameter of the gas. The analytical solutions for one dimensional adiabatic flow behind the strong imploding shock waves in non-ideal gas with a constant ratio of specific heats are derived by assuming that the disturbances due to the reflections, wave interactions in the wake, etc., do not overtake the imploding shock waves. The analytic expression for the propagation velocity of shock is obtained by substituting the generalized shock jump relations derived by the author [36] into the negative characteristic equation. Two cases are considered-cylindrical and spherical converging shocks to highlight the differences between the 2D and 3D convergence. The general non-dimensional forms of the analytical expressions for the distribution of pressure, density, particle velocity, speed of sound and adiabatic compressibility of the medium just behind imploding shock front are obtained, assuming the medium to be inviscid, non-heat conducting, electrically infinitely conducting, initially uniform and at rest. Most of the prior studies have remained focused on the propagation of shocks in an ideal or non-ideal gaseous media without discussing the change in entropy across the shock front. The expression for the change in entropy across the imploding shock front is also derived. The numerical estimations of flow variables behind the imploding front with cylindrical ($\alpha = 1$) and spherical shock ($\alpha = 2$) symmetries are carried out using MATHEMATICA and MATLAB codes. The effects of non-ideal gas are investigated on the flow field behind the shock as the imploding shock wave propagates towards the centre of convergence. This model appropriately makes obvious the effects due to an increase in (i) the propagation distance from the centre of convergence, (ii) the non-idealness parameter and (iii) the adiabatic index, on the propagation velocity of shock, pressure, density, particle velocity, speed of sound, compressibility of medium and the change in entropy across the shock front. The results are discussed by comparison with those for the case of a perfect gas flow. Thus, the results provided a clear picture of whether and how the non-idealness parameter and the adiabatic index affect the flow field just behind the imploding shock front.

    The rest of the paper is organized as follows: Section 2 describes the general assumptions and notations, equation of state for non-ideal gas and set of generalized jump relations. In Section 3 geometrical shock dynamics approach is used to obtain the analytical solutions. A brief discussion of the results is presented in Section 4. The

findings are concluded in Section 5 with details on which effects were accounted for and which were not.

## 2. Equations of motion and shock jump relations

The non-steady, one dimensional flow field in non-ideal gas is a function of two independent variables; the time $t$ and the space coordinate $r$. In order to get some essential features of shock wave propagation, it is assumed that the equilibrium-flow condition is maintained in the flow field. The conservation equations governing the flow of a one-dimensional, inviscid, non-ideal gas under an equilibrium condition can be expressed conveniently in Eularian coordinates as follows:

$$\frac{\partial u}{\partial t} + u\frac{\partial u}{\partial r} + \frac{1}{\rho}\frac{\partial p}{\partial r} = 0 \tag{1}$$

$$\frac{\partial \rho}{\partial t} + u\frac{\partial \rho}{\partial r} + \rho\left(\frac{\partial u}{\partial r} + j\frac{u}{r}\right) = 0 \tag{2}$$

$$\frac{\partial e}{\partial t} + u\frac{\partial e}{\partial r} - \frac{p}{\rho^2}\left(\frac{\partial \rho}{\partial t} + u\frac{\partial \rho}{\partial r}\right) = 0 \tag{3}$$

where $u(r,t)$ is the particle velocity, $\rho(r,t)$ the density, $p(r,t)$ the pressure, $e(r,t)$ the internal energy of non-ideal gas per unit mass, and $r$ is the distance from the origin, O. The geometrical factor $j$ is defined by $j = d\,\ln A/d\,\ln r$, where $A$ is the flow cross-section area [37]. Then the one-dimensional flow in plane, cylindrical and spherical symmetry is characterized by $j = 0, 1,$ and 2, respectively.

The equation of state of an ideal gas is generally applied to actual gases with sufficient accuracy. Obviously, this approximation is inadequate, and it is necessary to take account of the deviations of an actual gas from the ideal state which result from the interaction between its component molecules. A short description of the equation of state for non-ideal gas presented here was given in the recent paper of the author [36]. The information given in that paper is repeated here for completeness. The equation of state for a non-ideal gas is obtained by considering an expansion of the pressure $p$ in powers of the density $\rho$ as [35] $p = \Gamma \rho T[1+ \rho C_1(T) + \rho^2 C_2(T) + .....]$, where $\Gamma$ is the gas constant, $p$, $\rho$ and $T$ are the pressure, density, and temperature of the non-ideal gas, respectively, and $C_1(T), C_2(T)$, are virial coefficients. The first term in the expansion corresponds to an ideal gas. The second term is obtained by taking into account the interaction between pairs of molecules, and subsequent terms must involve the interactions between the groups of three, four, etc. molecules. In the high temperature range the coefficients $C_1(T)$ and $C_2(T)$ tend to constant values equal to $b$ and $(5/8)b^2$, respectively. For gases $b\rho \ll 1$, $b$ being the internal volume of the molecules, and therefore it is sufficient to consider the equation of state in the form [38]

$$p = \Gamma \rho T [1 + b\rho] \qquad (4)$$

In this equation the correction to pressure is missing due to the neglect of second and higher powers of $b\rho$ i.e. due to the neglect of interactions between groups of three, four, etc. molecules of the gas. Roberts and Wu [25, 39] have used an equivalent equation of state to study the shock theory of sonoluminescence. The internal energy $e$ per unit mass of the non-ideal gas is given as

$$e = p/\rho(\gamma - 1)(1 + b\rho), \qquad (5)$$

where $\gamma$ is the adiabatic index. Eq. (5) implies that

$$C_p - C_v = \Gamma(1 + b^2\rho^2/(1 + 2b\rho)) \cong \Gamma$$

neglecting the second and higher powers of $b\rho$. Here $C_p$ and $C_v$ are the specific heats of the gas at constant pressure and constant volume, respectively. The non-ideal gas effects can be expressed in the fundamental equations according to Chandrasekhar [40], by two thermodynamical variables, namely by the sound velocity factor (the isentropic exponent) $\Gamma^*$ and a factor $K$, which contains internal energy as follows:

$$\Gamma^* = (\partial \ln p / \partial \ln \rho)_S \text{ and } K = -\rho(\partial e/\partial \rho)_P / p$$

Using the first law of thermodynamics and the Eqs. (4) and (5), we obtain $\Gamma^* = \gamma(1 + 2b\rho)/(1 + b\rho)$ and $K = 1/(\gamma - 1)$, neglecting the second and higher powers of $b\rho$. This shows that the isentropic exponent $\Gamma^*$ is non-constant in the shocked gas, but the factor $K$ is constant for the simplified equation of state for non-ideal gas in the form given by Eq. (4).

The isentropic velocity of sound, $a$ in non-ideal gas is given by

$$a^2 = \Gamma^* p/\rho$$

The deviation of the behavior of a non-ideal gas from that of a perfect gas is indicated by the adiabatic compressibility defined as,

$$\tau = \frac{1}{\rho}\left(\frac{\partial \rho}{\partial p}\right)_S = \frac{(1 + b\rho)}{\gamma(1 + 2b\rho)p} \qquad (6)$$

where $(\partial \rho/\partial p)_S$ denotes the derivative of $\rho$ with respect to $p$ at the constant entropy $s$. We can write an explicit formula for the change in entropy $\Delta s$ across the shocks of arbitrary strength in non-ideal gas as

$$\Delta s = C_v \ln p - C_p \ln \rho - C_p b\rho \qquad (7)$$

Using Eq. (5), the Eq. (3) transforms into

$$\frac{\partial p}{\partial t} + u\frac{\partial p}{\partial r} + \rho a^2 \left(\frac{\partial u}{\partial r} + j\frac{u}{r}\right) = 0 \qquad (8)$$

where $a^2 = \gamma p[1+b\rho/(1+b\rho)]/\rho$ (9)

Let $p_o$ and $\rho_o$ denote the undisturbed values of pressure and density in-front of the shock wave and $u$, $p$ and $\rho$ be the values of the respective quantities at any point immediately after the passage of the shock, then $p$, $\rho$ and $u$ in terms of the undisturbed values of these quantities can be expressed by means of the following equations [36]

$$p = \frac{\rho_o a_o^2}{\delta} \left\{ \frac{2(\gamma(\delta^2-1)+\delta)M^2}{\gamma+1} - \frac{\gamma-1}{\gamma(\gamma+1)} \right\}$$

$$\rho = \frac{\rho_o[\gamma^2(4\delta^2-3)+2\gamma(2\delta-1)+1]M^2}{(\gamma+1)[(\gamma-1)M^2+2]}$$

$$u = \frac{2a_o(\gamma+1)}{[\gamma^2(4\delta^2-3)+2\gamma(2\delta-1)+1]} \left\{ \frac{M[2\gamma^2(\delta^2-1)+\gamma(2\delta-1)+1]}{\gamma+1} - \frac{1}{M} \right\}$$

$$U = a_o M$$

where $U$ is the shock velocity, $a_o (=\sqrt{\gamma \delta p_o/\rho_o})$ is the speed of sound in undisturbed medium, $\delta = (1+2b\rho_o)/(1+b\rho_o)$, $b\rho_o$ is the non-idealness parameter of the gas and, $M$ is Mach number.

For strong shocks, $U \gg a_o$, thus the pressure, density, particle velocity and sound speed just behind the strong shock can be, respectively, written as

$$p = \frac{2\rho_o[\gamma(\delta^2-1)+\delta]U^2}{\delta[2\gamma\delta^2-(\gamma-1)(2\delta-1)]} \tag{10}$$

$$\rho = \frac{\rho_o[\gamma^2(4\delta^2-3)+2\gamma(2\delta-1)+1]}{(\gamma-1)(\gamma+1)} \tag{11}$$

$$u = \frac{2[2\gamma^2(\delta^2-1)+\gamma(2\delta-1)+1]U}{\gamma^2(4\delta^2-3)+2\gamma(2\delta-1)+1} \tag{12}$$

$$a^2 = \frac{\gamma\delta(\gamma\delta-\gamma+1)U^2}{\gamma^2(4\delta^2-3)+2\gamma(2\delta-1)+1} \tag{13}$$

## 3. The geometrical shock dynamics theory and analytical solutions

In this section, we developed the geometrical shock dynamics model to provide a simplified, approximate treatment for the propagation of strong imploding shock waves in non-ideal fluids. The geometrical shock dynamics approach due to Whitham [34] provides practically accurate results especially for continuously accelerating imploding shocks. Consequently, the present model is well suited for studying shocks in the self-propagating limit, in which the front is moving steadily or accelerating.

According to the geometrical shock dynamics approach, the characteristic form of the governing Eqs. (1), (2) and (8), is easily obtained by forming a linear combination of Eqs. (1) and (8) in only one direction in $(r,t)$-plane. The linear combination of these two equations can be written as

$$\frac{\partial p}{\partial t} + (u+\lambda)\frac{\partial p}{\partial r} \pm \lambda\rho\frac{\partial u}{\partial t} + \rho(a^2+u\lambda)\frac{\partial u}{\partial r} + j\rho\, a^2\frac{u}{r} = 0 \qquad (14)$$

The conditions that this combination involves the derivatives in only one direction, are given by

$$\frac{\partial p}{\partial t} = (u+\lambda)\frac{\partial p}{\partial r} \quad \text{or} \quad \frac{\partial r}{\partial t} = (u+\lambda) \qquad (15)$$

and $\lambda\rho\dfrac{\partial u}{\partial t} = \rho(a^2+\lambda u)\dfrac{\partial u}{\partial r}$ or $\lambda\dfrac{\partial r}{\partial t} = a^2+\lambda u$ \qquad (16)

Eqs. (15) and (16) give $\lambda = \pm a$ i.e., $\dfrac{\partial r}{\partial t} = u \pm a$ \qquad (17)

It shows the fact that the characteristic curves in $(r,t)$-plane represent the motion of possible disturbances whose velocity differs from the velocity of non-ideal gas $u$ by the value $\pm a$ (speed of sound), respectively, for diverging and converging shock waves. Now, Eq. (14) can be written as

$$\frac{\partial p}{\partial t} + (u\pm a)\frac{\partial p}{\partial r} \pm a\rho\frac{\partial u}{\partial t} \pm \rho a\,(u\pm a)\frac{\partial u}{\partial r} + j\rho\, a^2\frac{u}{r} = 0 \qquad (18)$$

The above Eq. (18) is exact and holds throughout the flow since it is just a combination of the basic Eqs. (1), (2) and (8). By using above Eq. (18) we may write the characteristic form of the governing Eqs. (1), (2) and (8) i.e. the form in which equation contains derivatives in only one direction in the $(r,t)$-plane, as

$$dp + \rho\, a\, du + j\,\rho\, a^2\,\frac{u}{u+a}\frac{dr}{r} = 0 \quad \text{along } C_+ \text{ i.e. } \frac{dr}{dt} = u+a \qquad (19)$$

and

$$dp - \rho\, a\, du + j\,\rho\, a^2\,\frac{u}{u-a}\frac{dr}{r} = 0 \quad \text{along } C_- \text{ i.e. } \frac{dr}{dt} = u-a \qquad (20)$$

The Eqs. (19) and (20) represent the characteristic equations for exploding and imploding shock waves, respectively. The geometrical shock dynamics approach states that when relevant equations are written first in the characteristics form, the differential relation which must be satisfied along a characteristic can be applied to the flow quantities just behind the shock front. Together with the shock jump relations, this rule determines the propagation of the shock waves. We assume here that the shock jump relations to hold, of course, within the order of approximation determine by a constant value of $U$. We apply here the differential relation (20) along the negative characteristic $C_-$ behind the shock wave. Together with the shock jump relations, we are able to describe the shock velocity $U$ or the related quantities in terms of the quantities just ahead of the shock front. Eq. (20) is valid only along the negative characteristic curve $C_-$ in the $(r,t)$-plane, behind the imploding shock front.

The idea of the characteristic rule of Whitham [34] is to apply, on the negative characteristic curve $C_-$ along the imploding shock front. We thus neglect the difference in the constants of integration obtained when Eq. (20) is solved on different characteristics that intersect the shock front. These differences arise from the non-uniformity of the flow behind the shock, so the characteristic rule effectively ignores the influence of the flow behind the shock wave on the shock propagation. Because the effect of the flow behind the shock on the shock dynamics is ignored, the method is very good for situations where the shock wave accelerates with time, so that features of the flow behind do not 'catch up' with the shock. The excellent examples of flows with this characteristic are converging shock waves, which are the subject of this paper. For spherically and cylindrically symmetric implosions in ideal gases, the results of the Whitham's rule can be compared with the exact solutions and are correct to three significant figures.

Now, assuming that the negative characteristic curve $C_-$ applies on the shock front, we can use the shock jump relations given by Eqs. (10) - (13) to write the quantities in it, which are those immediately behind the shock, in terms of those ahead of the shock and the shock velocity. The shock jump relations we use here are the shock conditions for the non-ideal gas [36] rather than the ideal gas shock conditions used by Whitham [34].

Now, substituting the shock jump relations given by Eqs. (10) - (13) into the negative characteristic curve $C_-$ we get a first order ordinary differential equation in $U^2$ as

$$\frac{dU^2}{U^2} + 2j\mathrm{N}\,\frac{dr}{r} = 0 \qquad (21)$$

where $\mathrm{N} = \dfrac{\mathrm{N}_2\mathrm{N}_3\mathrm{N}_5}{2(\gamma-1)(\gamma+1)(\mathrm{N}_3 - \mathrm{N}_2\sqrt{\mathrm{N}_5})}\left[\dfrac{2\mathrm{N}_1}{\mathrm{N}_4} - \dfrac{\mathrm{N}_3\sqrt{\mathrm{N}_5}}{2(\gamma-1)(\gamma+1)}\right]^{-1}$

$\mathrm{N}_1 = \gamma[\gamma(\delta^2 - 1) + \delta]$, $\mathrm{N}_2 = \gamma^2(4\delta^2 - 3) + 2\gamma(2\delta - 1) + 1$,

$\mathrm{N}_3 = \gamma^2(4\delta^2 - 4) + 2\gamma(2\delta - 1) + 2$, $\mathrm{N}_4 = \delta\gamma[2\gamma\delta^2 - (\gamma-1)(2\delta - 1)]$,

$\mathrm{N}_5 = \gamma\delta(\gamma\delta - \gamma + 1)/[\gamma^2(4\delta^2 - 3) + 2\gamma(2\delta - 1) + 1]$, $\delta = (1 + 2b\rho_o)/(1 + b\rho_o)$

On integration, Eq. (21) yields, the square of the shock velocity as $U^2 = K' r^{-2j\mathrm{N}}$, where $K'$ is the constant of integration. The analytical expression for the non-dimensional propagation velocity of shock may be written as

$$\frac{U}{a_o} = \frac{K}{\sqrt{\gamma\delta}} r^{-j\mathrm{N}} \qquad (22)$$

where the constant $K = \sqrt{K'\rho_o/p_o}$. This equation, valid for the imploding shocks propagating in non-ideal gas, is the main result of the present investigation. Thus, the characteristic rule for the propagation velocity is $U \propto r^{-N}$ for cylindrical shock waves and is $U \propto r^{-2N}$ for spherical shock waves, where $r$ is the radius of the shock front from the centre of convergence. For infinitely strong shock waves in ideal gas, Whitham [34, p273] obtained the characteristic rule as $U \propto r^{-N}$ for cylindrical shocks and $U \propto r^{-2N}$ for spherical shocks, where $N = [1 + 2/\gamma + \sqrt{2\gamma/(\gamma-1)}]^{-1}$. The characteristic rule is used for the problem of a shock propagating through a uniform and non-uniform density media. It is also the basis for the geometrical treatment of two and three dimensional shock propagation. Now, the corresponding analytical expressions for the distribution of pressure $p$, density $\rho$, particle velocity $u$, and sound speed $a$ just behind the imploding shock front can be easily written as

$$\frac{p}{p_o} = \frac{2K^2[\gamma(\delta^2-1)+\delta]}{\delta[2\gamma\delta^2-(\gamma-1)(2\delta-1)]} r^{-2jN} \tag{23}$$

$$\frac{\rho}{\rho_o} = \frac{[\gamma^2(4\delta^2-3)+2\gamma(2\delta-1)+1]}{(\gamma-1)(\gamma+1)} \tag{24}$$

$$\frac{u}{a_o} = \frac{2K[2\gamma^2(\delta^2-1)+\gamma(2\delta-1)+1]}{\sqrt{\gamma\delta}[\gamma^2(4\delta^2-3)+2\gamma(2\delta-1)+1]} r^{-jN} \tag{25}$$

$$\frac{a}{a_o} = \left\{ \frac{K^2(\gamma\delta-\gamma+1)}{\gamma^2(4\delta^2-3)+2\gamma(2\delta-1)+1} \right\}^{1/2} r^{-jN} \tag{26}$$

Using Eqs. (6), (24) and (26), the adiabatic compressibility of the medium may be expressed as

$$\tau(p_o) = \frac{(\gamma-1)(\gamma+1)}{K^2\gamma\delta(\gamma\delta-\gamma+1)} r^{2jN} \tag{27}$$

The analytical expression for the change in entropy across a strong imploding shock front in non-ideal gas flow is easily obtained by using the equations (7), (23) and (24) as

$$\begin{aligned}\frac{\Delta s}{\Gamma} &= \frac{1}{(\gamma-1)} \ln\left[\frac{2K^2[\gamma(\delta^2-1)+\delta]r^{-2jN}}{\delta[2\gamma\delta^2-(\gamma-1)(2\delta-1)]}\right] \\ &\quad - \frac{\gamma}{(\gamma-1)} \ln\left[\frac{\gamma^2(4\delta^2-3)+2\gamma(2\delta-1)+1}{(\gamma-1)(\gamma+1)}\right] \\ &\quad - \frac{b\rho_o\gamma[\gamma^2(4\delta^2-3)+2\gamma(2\delta-1)+1]}{(\gamma-1)^2(\gamma+1)} \end{aligned} \tag{28}$$

Thus, the influence of non-ideal gas on the propagation velocity of imploding shock wave and the flow field parameters behind the imploding shock front can be explored from the above analytical expressions (22)-(28).

## 4. Results and Discussion

In the present paper the general analytical solution for strong imploding shock waves in non-ideal gas was obtained by adopting the geometrical shock dynamics, due to Whitham [34] and further the general solution was examined and explored for the cylindrical and spherical shock waves. The goal of the present investigation was to examine the effects due to the non-ideal gas on the imploding shock waves as they focus at the centre of convergence and the region of flow field immediately behind the shock front. The equation of state for non-ideal gas was considered as given by Landau and Lifshitz [35]. The general characteristic rule for the propagation velocity $U/a_o$ of strong imploding shock waves in non-ideal gas is given by Eq. (22). The general non-dimensional forms of the analytical expressions for the distribution of the pressure $p/p_o$, the density $\rho/\rho_o$, the particle velocity $u/a_o$, the speed of sound $a/a_o$, and the adiabatic compressibility $\tau(p_o)$ of medium behind the imploding shock front are given by Eqs. (23) – (27), respectively. The change in entropy $\Delta s/\Gamma$ across the imploding shock front in the non-ideal gas is given by the Eq. (28). These analytical expressions were derived by assuming that the disturbances due to the reflections, wave interactions in the wake, etc., do not overtake the imploding shock waves. It is worth mentioning that the effects due to the non-ideal gas enter through the parameter of non-idealness $b\rho_o$. It is notable that the analytical solutions for the cylindrical and spherical imploding shock waves are principally identical and only a geometrical factor $j$ differs the two cases.

The general non-dimensional analytical expressions for the propagation velocity of shock, the pressure, the density, the particle velocity, the sound speed, the adiabatic compressibility of medium and the change in entropy across the shock front are the functions of the propagation distance $r$ from the origin O, the adiabatic index $\gamma$ and the non-idealness parameter $b\rho_o$ of the gas. Therefore, the values of the constant parameters are taken to be $b\rho_o = 0$, 0.03125, 0.06250, 0.09375, 0.12500, 0.15625, 0.18750, 0.21875, 0.25000; $\gamma = 6/5, 11/9, 5/4, 9/7, 4/3, 7/5, 3/2, 5/3, 2$ for the general purpose of numerical computations. The planar case $j = 0$ is not of interest since no area convergence and shock amplification exist and it is simply the ordinary planar blast wave problem. The value $b\rho_o = 0$, corresponds to the case of a perfect gas. It is very useful to mention that the present analysis serves an analytical description for the propagation of strong imploding shock waves through an in-viscid, non-heat conducting and electrically infinitely conducting real gases.

The N-parameter is a function of non-idealness parameter $b\rho_o$ and adiabatic index $\gamma$ of the gas. The variation of N-parameter with non-idealness parameter $b\rho_o$ and adiabatic index $\gamma$ are shown through Table1. It is important to note that the N-parameter increases linearly with increase in the value of non-idealness parameter $b\rho_o$ whereas it decreases exponentially with increase in the value of adiabatic index $\gamma$.

**Table 1** Variation of N-parameter with non-idealness parameter $b\rho_o$ and adiabatic index $\gamma$.

| $b\rho_o$ | $\gamma=6/5$ | $\gamma=11/9$ | $\gamma=5/4$ | $\gamma=9/7$ | $\gamma=4/3$ | $\gamma=7/5$ | $\gamma=3/2$ | $\gamma=5/3$ | $\gamma=2$ |
|---|---|---|---|---|---|---|---|---|---|
| 0 | 10.2473 | 8.68496 | 7.24771 | 5.93570 | 4.74914 | 3.68849 | 2.75491 | 1.95192 | 1.29521 |
| 0.03125 | 11.3858 | 9.63735 | 8.02957 | 6.56232 | 5.23548 | 4.04886 | 3.00217 | 2.09495 | 1.32676 |
| 0.06250 | 12.5654 | 10.6264 | 8.84420 | 7.21849 | 5.74904 | 4.43534 | 3.27649 | 2.27059 | 1.41284 |
| 0.09375 | 13.7762 | 11.6429 | 9.68283 | 7.89583 | 6.28149 | 4.83915 | 3.56761 | 2.46435 | 1.52313 |
| 0.12500 | 15.0104 | 12.6798 | 10.5393 | 8.58867 | 6.82751 | 5.25506 | 3.86994 | 2.66936 | 1.64656 |
| 0.15625 | 16.2621 | 13.7318 | 11.4089 | 9.29286 | 7.38337 | 5.67958 | 4.18005 | 2.88181 | 1.77795 |
| 0.18750 | 17.5262 | 14.7947 | 12.2878 | 10.0051 | 7.94620 | 6.11019 | 4.49558 | 3.09935 | 1.91452 |
| 0.21875 | 18.7986 | 15.8649 | 13.1731 | 10.7229 | 8.51378 | 6.54495 | 4.81483 | 3.32036 | 2.05453 |
| 0.25000 | 20.0757 | 16.9391 | 14.0620 | 11.4438 | 9.08424 | 6.98231 | 5.13649 | 3.54368 | 2.19688 |

4.1 Cylindrical shock waves: The non-dimensional analytical expressions for the shock velocity $U/a_o$, the pressure $p/p_o$, the density $\rho/\rho_o$, the particle velocity $u/a_o$, the sound speed $a/a_o$, the adiabatic compressibility $\tau(p_o)$ of the region just behind the shock and the change in entropy $\Delta s/\Gamma$ across the imploding cylindrical shock front propagating in the non-ideal gas are obtained by taking $j=1$ in the equations (22) - (28), respectively. In numerical computations, the value of constant $K=0.0926773$ which is obtained by assuming that the shock velocity, i.e. $U=5a_o$ at $r=0.2$ for $\gamma=5/3$ and $b\rho_o=0.125$. The variations of the shock velocity, pressure, particle velocity, sound speed, adiabatic compressibility and change in entropy with propagation distance $r$ for $\gamma=5/3$ and various values of non-idealness parameter $b\rho_o$ are shown in Fig. 1(a-f). It is remarkable that the shock velocity, pressure, particle velocity and speed of sound increase as the cylindrical shock wave approaches the centre of convergence (see, Fig. 1(a-d)). The compressibility of the medium behind the shock front decreases as the shock moves towards the centre of convergence (see, Fig. 1(e)) whereas the change in entropy across the front increases (see, Fig. 1(f)). It is notable that the distribution of density immediately behind the cylindrical shock front is independent of propagation distance $r$ (vide, Eq. (24)). Table 2 displays the variation of density with non-idealness parameter $b\rho_o$ and adiabatic index $\gamma$ of the gas. The density just behind the cylindrical shock front increases with increase in the value of non-idealness parameter $b\rho_o$ whereas it decreases with increase in the value of adiabatic index $\gamma$. It is remarkable that the density behind the cylindrical shock front remains unchanged with the propagation distance $r$.

**Table 2** Variation of density $\rho/\rho_o$ behind shock front with non-idealness parameter $b\rho_o$ and adiabatic index $\gamma$.

| $b\rho_o$ | $\gamma=6/5$ | $\gamma=11/9$ | $\gamma=5/4$ | $\gamma=9/7$ | $\gamma=4/3$ | $\gamma=7/5$ | $\gamma=3/2$ | $\gamma=5/3$ | $\gamma=2$ |
|---|---|---|---|---|---|---|---|---|---|
| 0 | 11.0000 | 10.0000 | 9.00000 | 8.00000 | 7.00000 | 6.00000 | 5.00000 | 4.00000 | 3.00000 |
| 0.03125 | 12.1360 | 11.0444 | 9.95296 | 8.86157 | 7.77030 | 6.67922 | 5.58843 | 4.49816 | 3.40894 |
| 0.06250 | 13.2271 | 12.0478 | 10.8685 | 9.68945 | 8.51063 | 7.33218 | 6.15433 | 4.97751 | 3.80277 |
| 0.09375 | 14.2754 | 13.0118 | 11.7483 | 10.4851 | 9.22227 | 7.96000 | 6.69861 | 5.43878 | 4.18204 |
| 0.12500 | 15.2828 | 13.9383 | 12.5940 | 11.2500 | 9.90653 | 8.56379 | 7.22222 | 5.88272 | 4.54733 |
| 0.15625 | 16.2513 | 14.8291 | 13.4071 | 11.9856 | 10.5646 | 9.14463 | 7.72608 | 6.31008 | 4.89920 |
| 0.18750 | 17.1828 | 15.6859 | 14.1893 | 12.6932 | 11.1979 | 9.70360 | 8.21108 | 6.72161 | 5.23823 |
| 0.21875 | 18.0791 | 16.5103 | 14.9420 | 13.3743 | 11.8074 | 10.2417 | 8.67811 | 7.11801 | 5.56498 |
| 0.25000 | 18.9418 | 17.3040 | 15.6667 | 14.0300 | 12.3943 | 10.7600 | 9.12800 | 7.50000 | 5.88000 |

Fig. 2(a-f) shows the variations of the shock velocity, pressure, particle velocity, sound speed, adiabatic compressibility and change in entropy with non-idealness parameter $b\rho_o$ for $r = 0.7$ and various values of adiabatic index $\gamma$ of the gas. It is notable that the shock velocity increases with increase in the value of non-idealness parameter $b\rho_o$ whereas it decreases with increase in the value of adiabatic index $\gamma$ (see, Fig. 2(a)). The pressure, particle velocity and sound speed just behind the shock front increase with increase in the value of non-idealness parameter $b\rho_o$ whereas these flow variables decrease with increase in the value of adiabatic index $\gamma$ (see, Fig. 2(b-d)). The adiabatic compressibility of medium after the passage of shock decreases with increase in the value of non-idealness parameter $b\rho_o$ whereas it increases with an increase in the value of adiabatic index $\gamma$ (see, Fig. 2(e)). The change in entropy across the shock front decreases with increase in the value of non-idealness parameter $b\rho_o$ however it increases with increase in the value of adiabatic index $\gamma$ of the gas (see, Fig. 2(f)). Thus, the Fig. 2(a-f) reveals clearly the effects of non-idealness parameter $b\rho_o$ and adiabatic index $\gamma$ of the gas on the shock velocity, the flow quantities just behind the shock front and the change in entropy across the cylindrical shock front. It is obvious that the effects due to the non-idealness parameter $b\rho_o$ and the adiabatic index $\gamma$ modify the numerical values of the flow variables from their values for the ideal gas case ($b\rho_o = 0$), but their trends of variations approximately remain unaffected, in general. For fully ionized medium $\gamma = 5/3$ and thus, the above results obtained for the strong imploding cylindrical shock waves are directly applicable to the stellar medium.

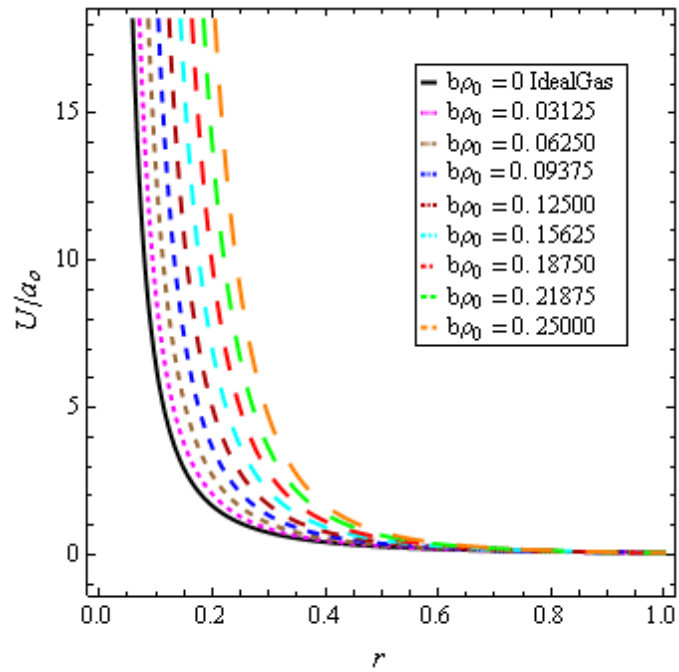

(a)

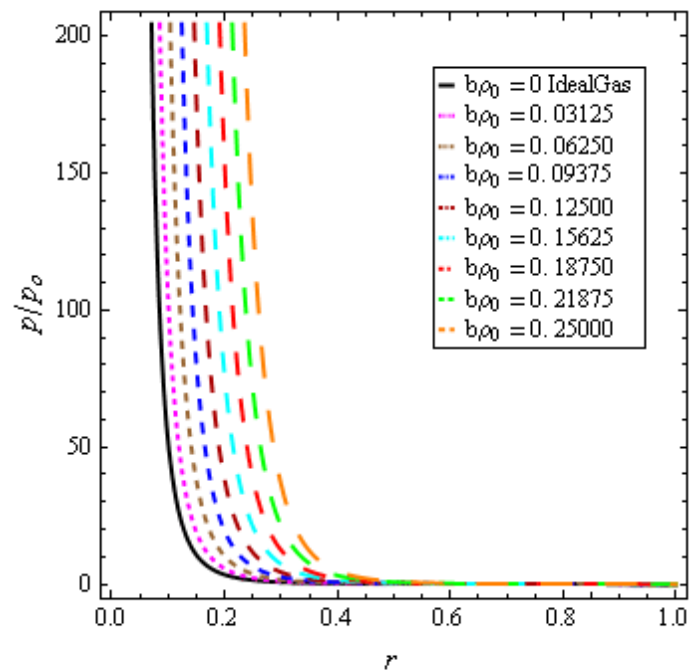

(b)

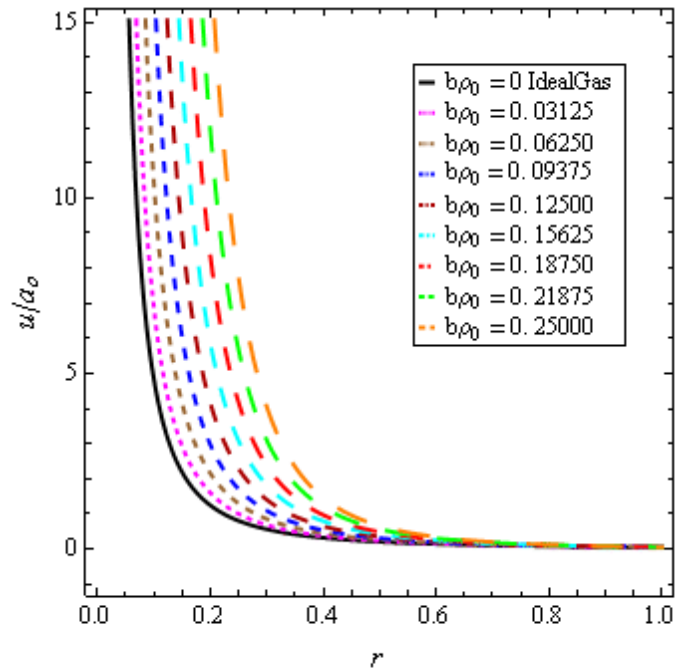

(c)

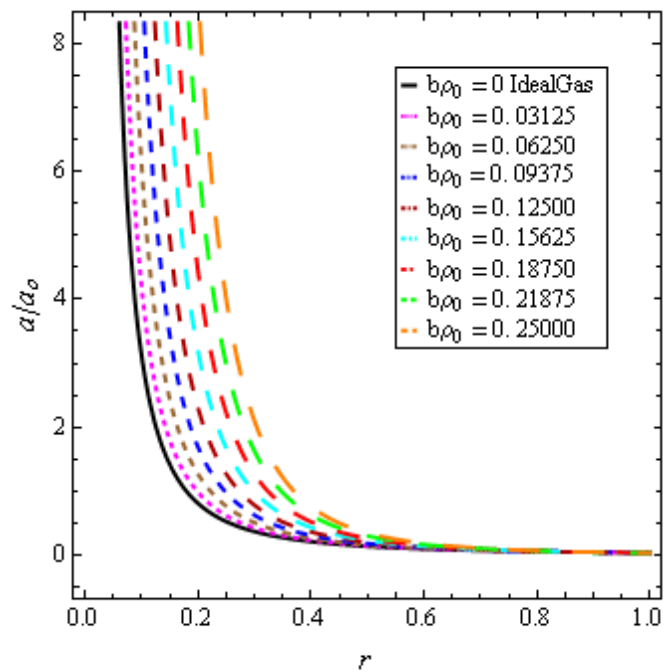

(d)

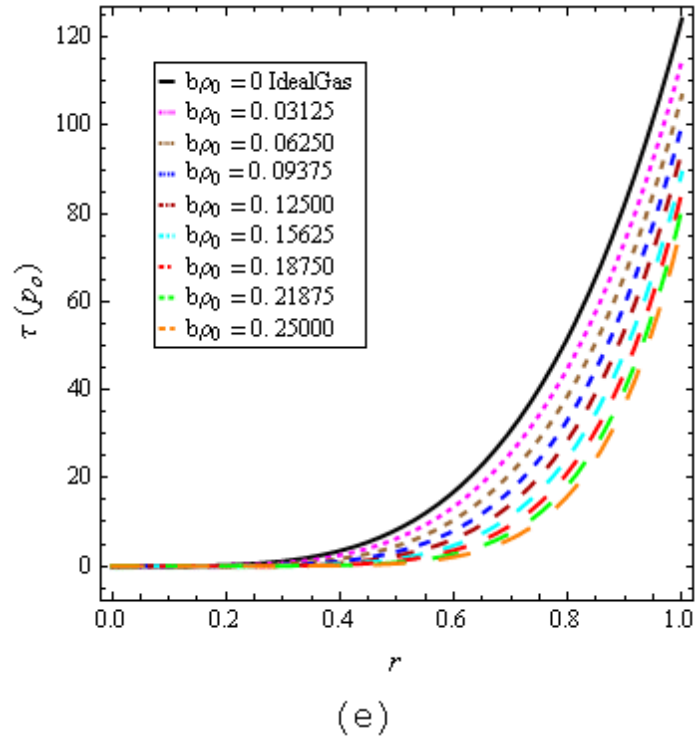

(e)

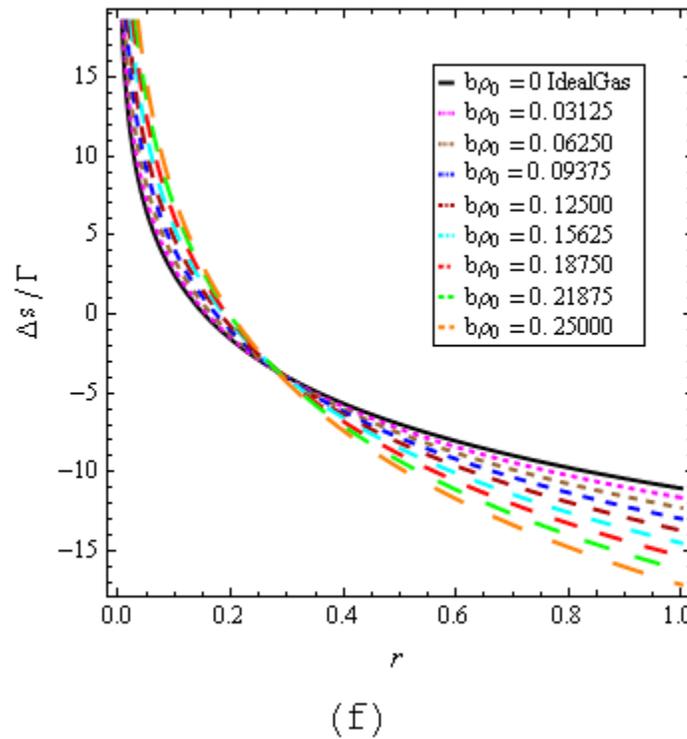

(f)

**Fig. 1.** The variations of (a) shock velocity, (b) pressure, (c) particle velocity, (d) sound speed, (e) adiabatic compressibility and (f) change-in-entropy distribution behind the strong imploding cylindrical shock wave with propagation distance.

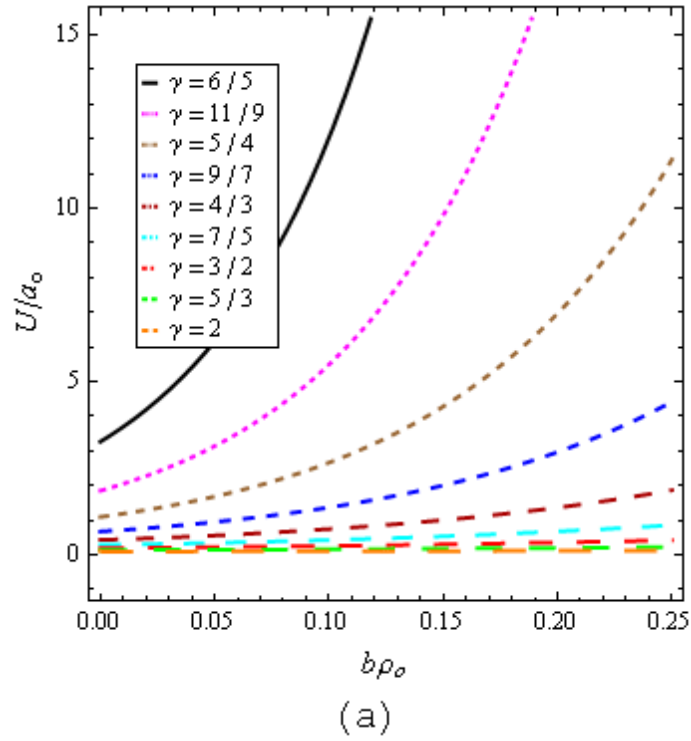

(a)

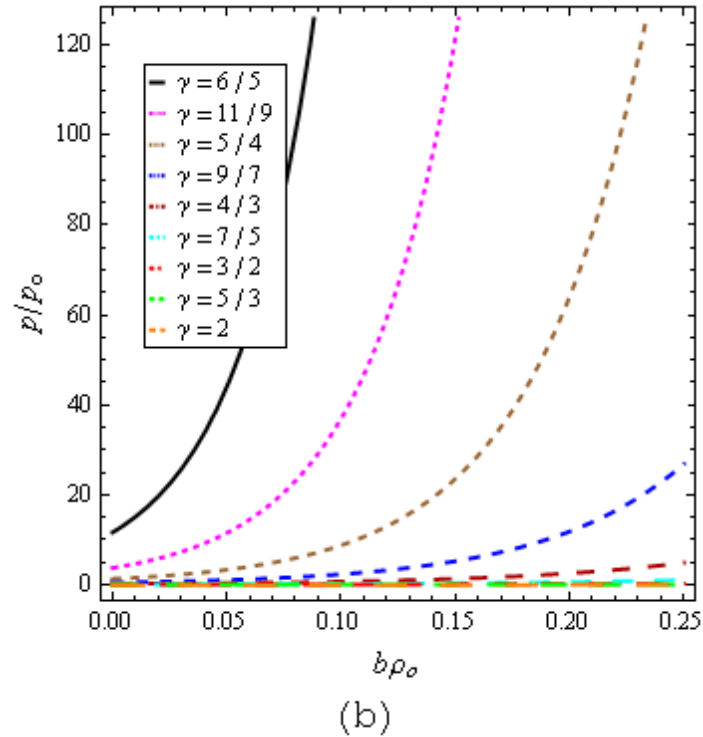

(b)

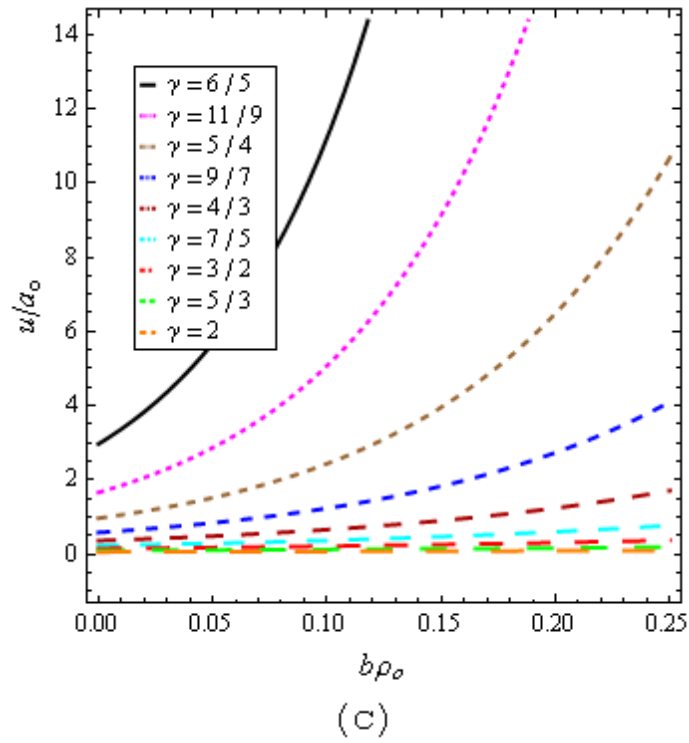

(c)

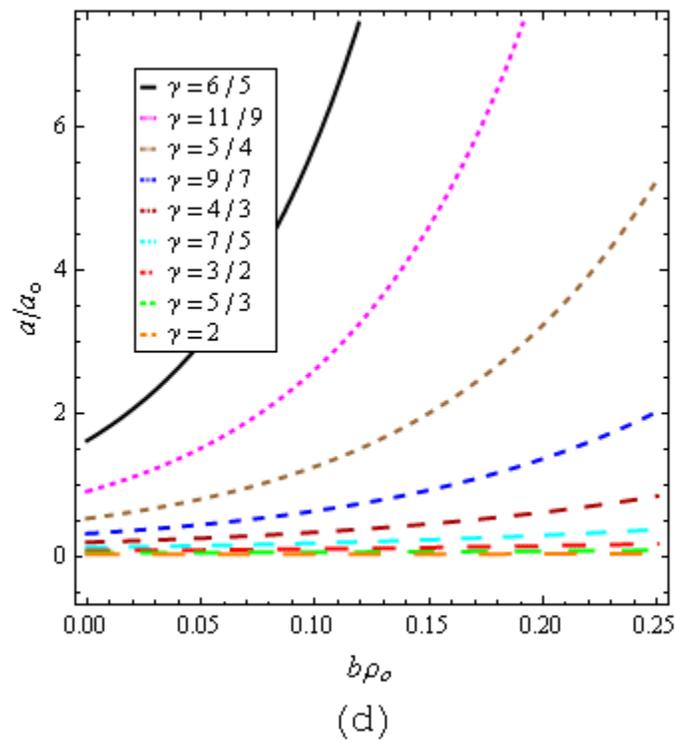

(d)

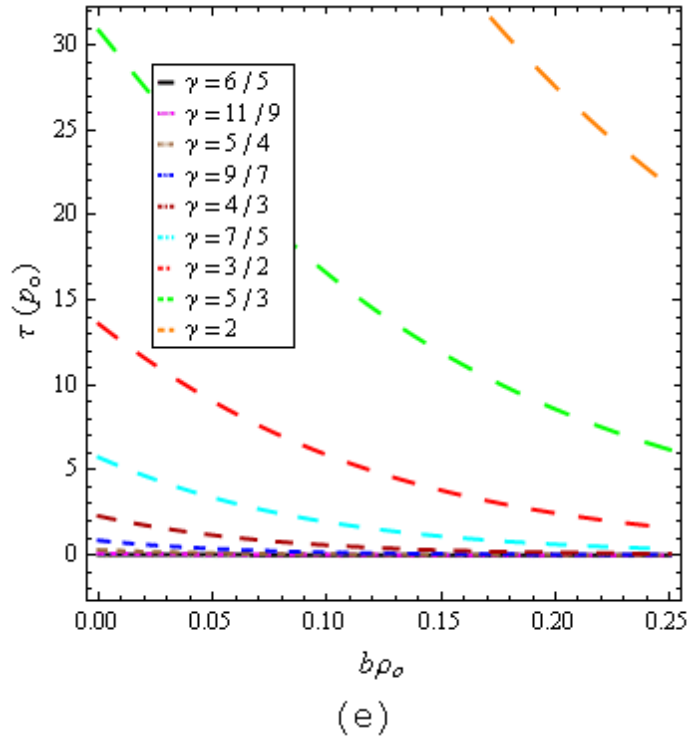

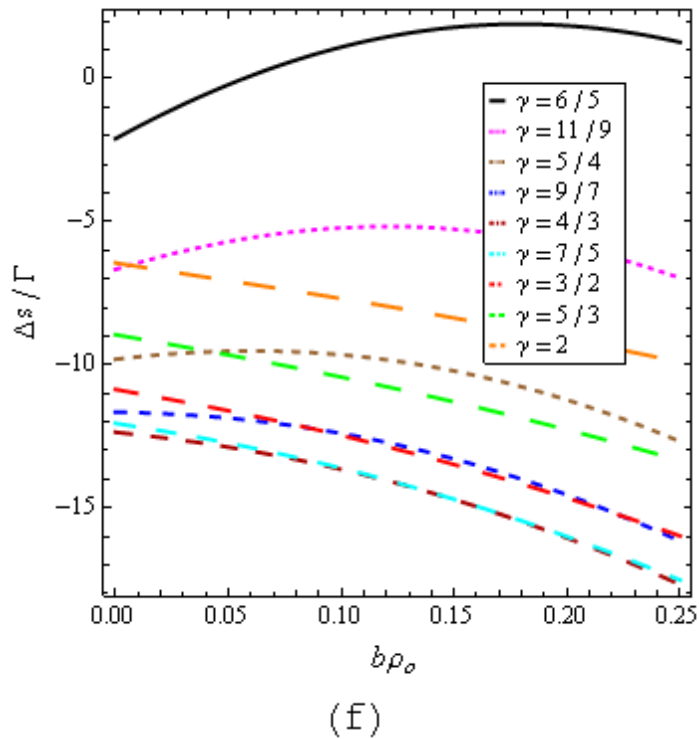

**Fig. 2.** The variations of (a) shock velocity, (b) pressure, (c) particle velocity, (d) sound speed, (e) adiabatic compressibility and (f) change-in-entropy distribution behind the strong imploding cylindrical shock wave with non-idealness parameter of the gas.

4.2 Spherical shock waves: The non-dimensional analytical expressions for the shock velocity $U/a_o$, the pressure $p/p_o$, the density $\rho/\rho_o$, the particle velocity $u/a_o$, the sound speed $a/a_o$, the adiabatic compressibility $\tau(p_o)$ of the region just behind the shock and the change in entropy $\Delta s/\Gamma$ across the strong imploding spherical shock front propagating in the non-ideal gas are obtained by taking $j = 2$ in Eqs. (22) - (28), respectively. In numerical computations, the value of constant $K = 0.00126233$ which is obtained by assuming that the Mach number, i.e. $U = 5a_o$ at $r = 0.2$ for $\gamma = 5/3$ and $b\rho_o = 0.125$. The variations of the shock velocity, pressure, particle velocity, sound speed, adiabatic compressibility and change in entropy with propagation distance $r$ for $\gamma = 5/3$ and different values of non-idealness parameter $b\rho_o$ of the gas are shown in Fig. 3(a-f). It is noteworthy that the shock velocity, pressure, particle velocity and sound speed increase as the spherical shock wave propagates towards the centre of convergence (see, Fig. 3(a-d)). The adiabatic compressibility of the medium just behind the shock decreases as the shock wave converges at the centre of convergence (see, Fig. 3(e)) however the change in entropy across the shock front increases (see, Fig. 3(f)). It is notable that the distribution of density immediately behind the spherical shock front is independent of the propagation distance $r$ (vide Eq. (24)). Table 2 displays the variation of density with non-idealness parameter $b\rho_o$ and adiabatic index $\gamma$ of the gas. The density just behind the spherical shock front increases with increase in the value of non-idealness parameter $b\rho_o$ of the gas whereas it decreases with increase in the value of adiabatic index $\gamma$. It is remarkable that the density behind the spherical shock front remains unchanged with the propagation distance $r$. Fig. 4(a-f) shows the variations of the shock velocity, pressure, particle velocity, sound speed, adiabatic compressibility and change in entropy with non-idealness parameter $b\rho_o$ of the gas for $r = 0.7$ and different values of the adiabatic index $\gamma$. It is worth mentioning that the shock velocity increases with increase in the value of non-idealness parameter $b\rho_o$ whereas it decreases with increase in the value of adiabatic index $\gamma$ of the gas (see, Fig. 4(a)). The pressure, particle velocity and sound speed just behind the shock front increase with increase in the value of non-idealness parameter $b\rho_o$ of the gas whereas these flow variables decrease with increase in the value of adiabatic index $\gamma$ of the gas (see, Fig. 4(b-d)). The adiabatic compressibility of the medium after the passage of shock decreases with increase in the value of non-idealness parameter $b\rho_o$ of the gas whereas it increases with an increase in the value of adiabatic index $\gamma$ of the gas (see, Fig. 4(e)). The change in entropy across the spherical shock front increases with increase in the value of non-idealness parameter $b\rho_o$ of the gas however it decreases with increase in the value of adiabatic index $\gamma$ of the gas (see, Fig. 4(f)). Thus, the Fig. 4(a-f) discloses clearly the effects due to the non-idealness parameter $b\rho_o$ and the adiabatic index $\gamma$ of the gas on the shock velocity, the flow quantities behind the spherical shock front and the change in entropy across the spherical shock front. Obviously, the effects due to the non-idealness parameter $b\rho_o$ of the gas and the adiabatic index $\gamma$ alter the numerical values of the flow quantities from their values for the ideal gas case ($b\rho_o = 0$), but their trends of variations approximately remain unchanged, in general. In case of fully ionized medium $\gamma = 5/3$ and thus, the above results obtained for the strong imploding spherical shock waves can also be directly applied to the imploding spherical shock waves propagating in the astrophysical fluids.

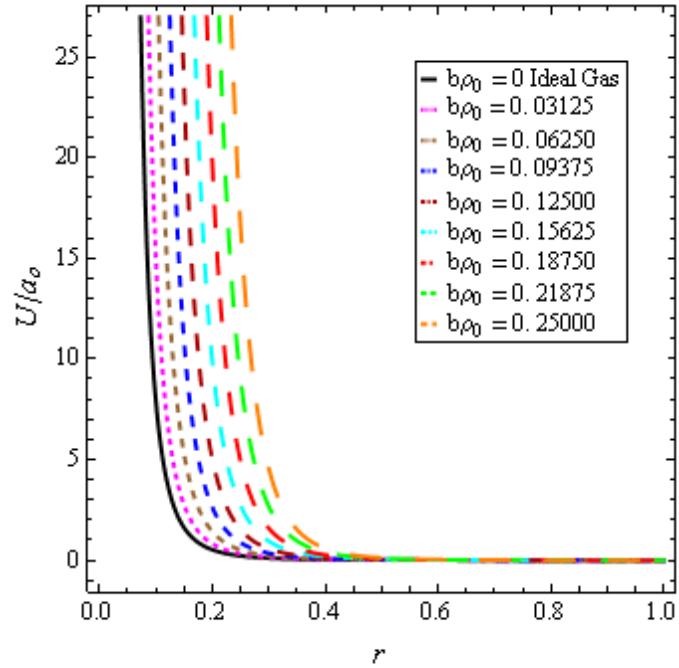

(a)

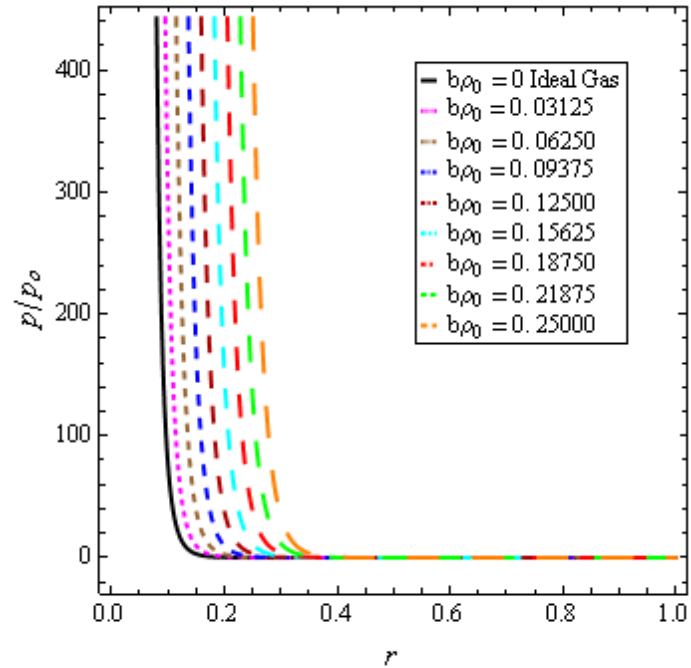

(b)

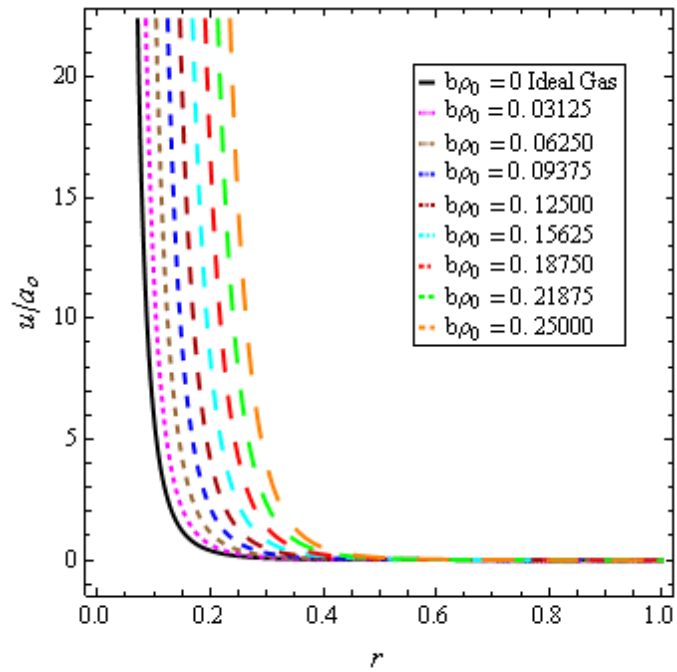

(c)

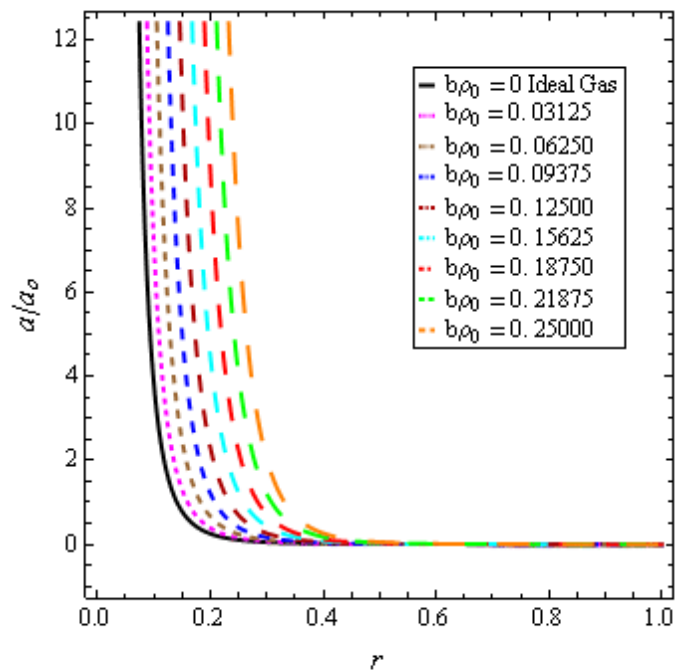

(d)

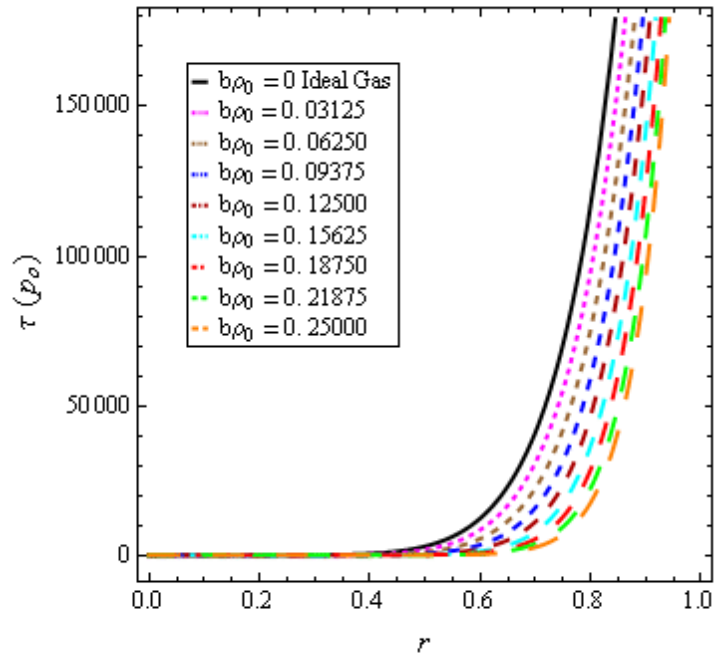

(e)

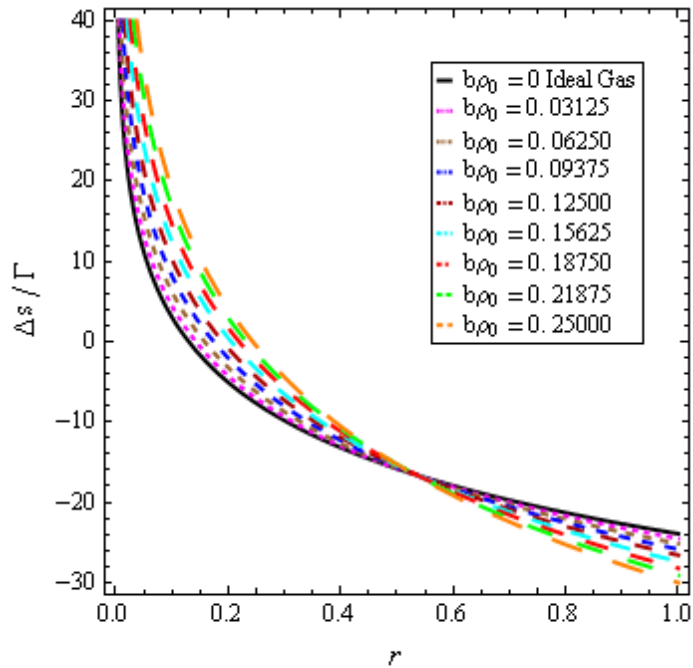

(f)

**Fig. 3.** The variations of (a) shock velocity, (b) pressure, (c) particle velocity, (d) sound speed, (e) adiabatic compressibility and (f) change-in-entropy distribution behind the strong imploding spherical shock wave with propagation distance.

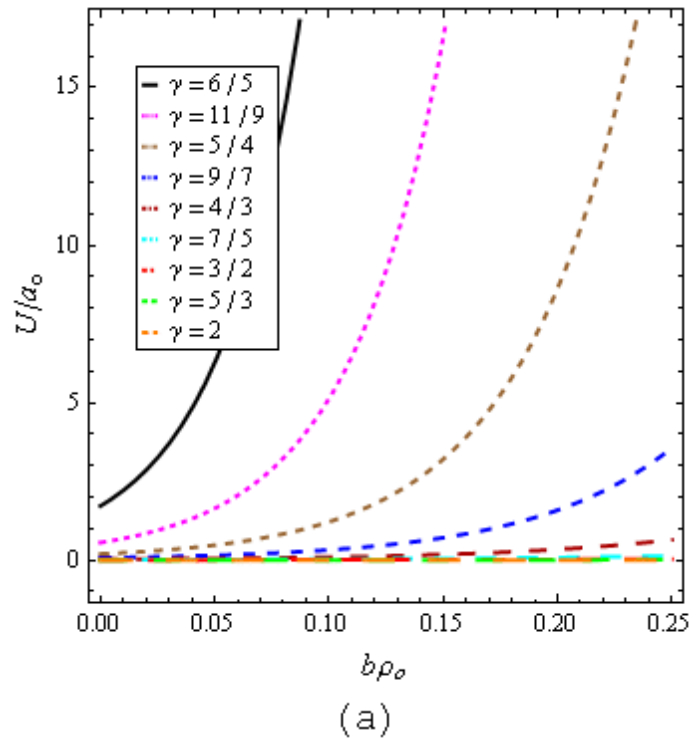

(a)

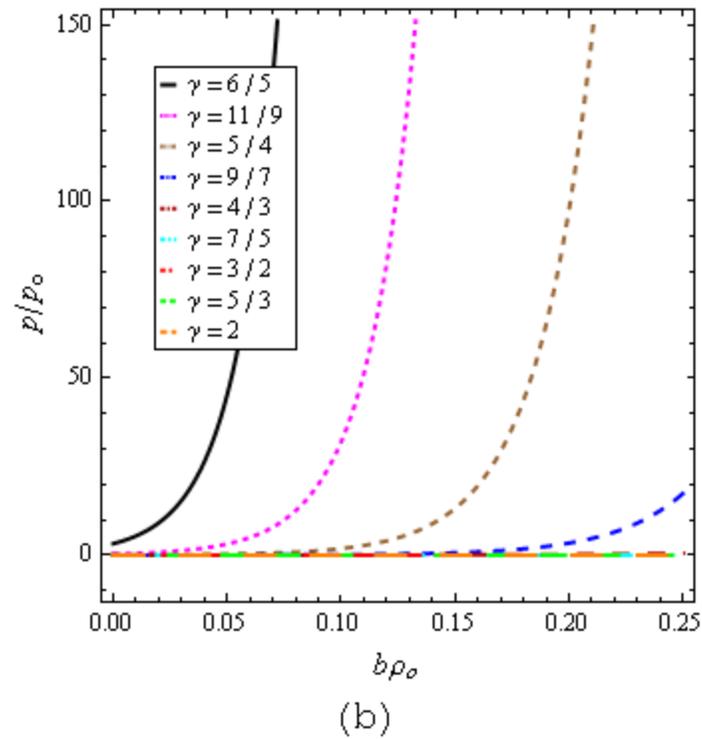

(b)

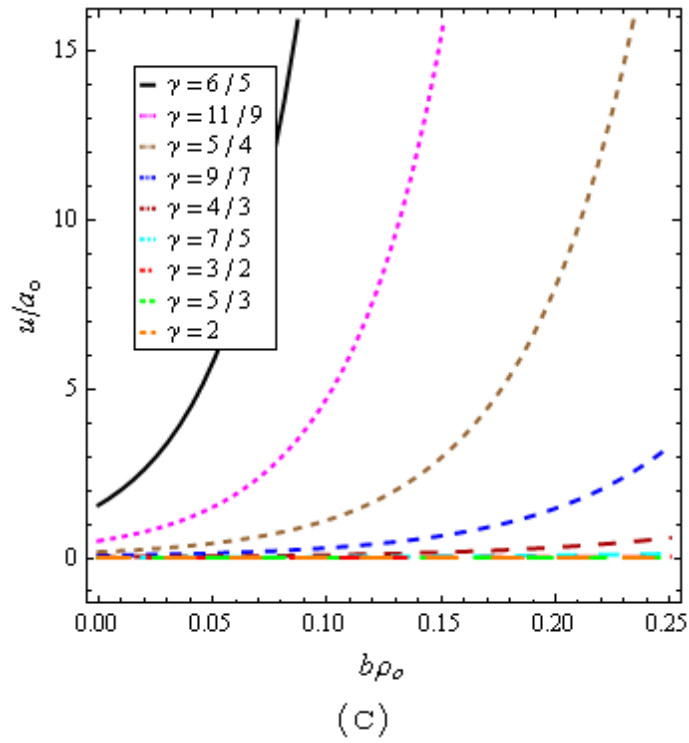

(c)

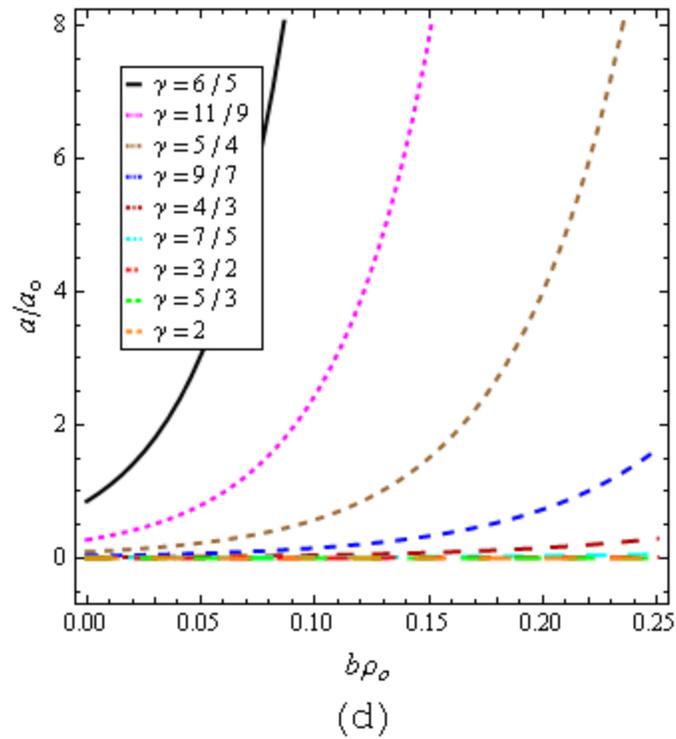

(d)

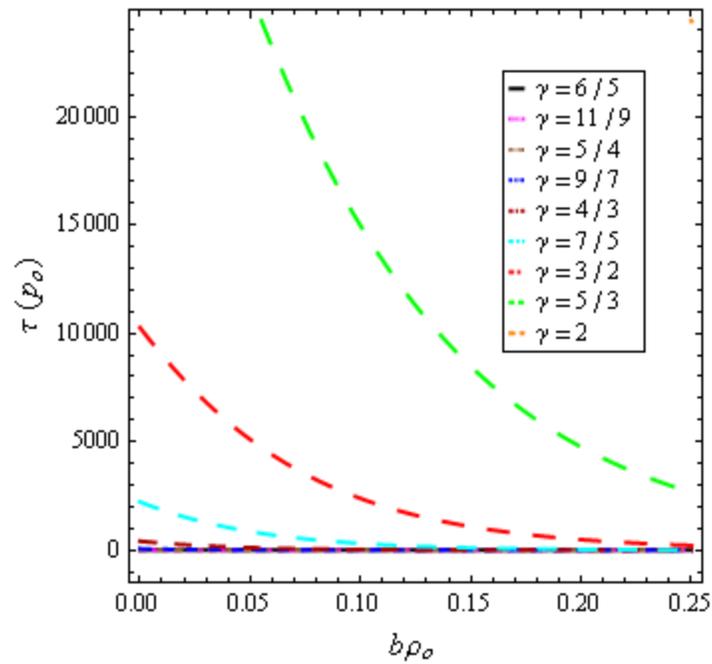

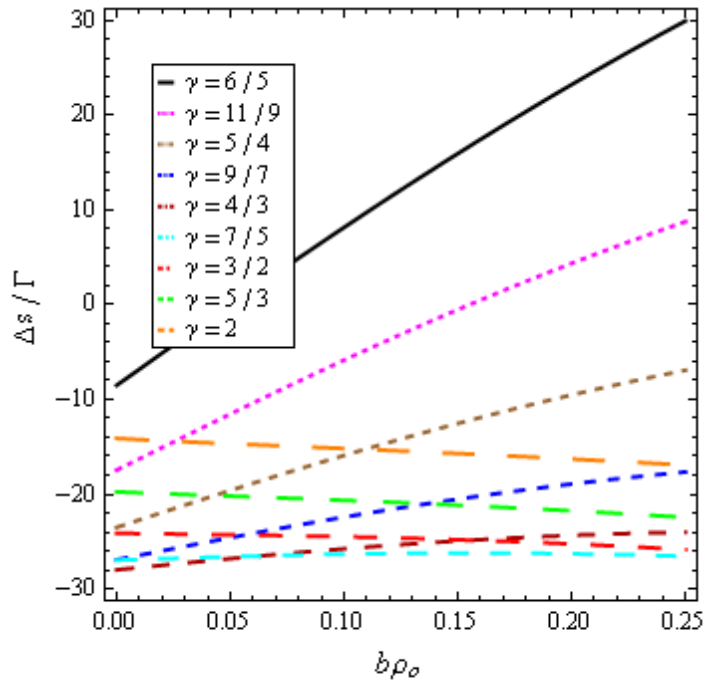

**Fig. 4.** The variations of (a) shock velocity, (b) pressure, (c) particle velocity, (d) sound speed, (e) adiabatic compressibility and (f) change-in-entropy distribution behind the strong imploding spherical shock wave with non-idealness parameter of the gas.

## 5. Conclusions

The investigations made in the present paper are intended to contribute to the understanding of the strong imploding shock waves in real gases, astrophysical fluids, etc. by giving, for the first time, the full analytical solutions for the flow field behind the imploding shock front. The analytical solutions were obtained using the geometrical shock dynamics approach [34]. The effects due to the non-ideal gas on the strong imploding cylindrical and spherical shock waves were studied in view of the equation of state as given by Landau and Lifshitz [35]. Such problem is of great interest in astrophysics and space science as it is highly relevant to the problem of the origin of cosmic rays [6, 7, 9, 12].

The following conclusions may be drawn from the findings of the current analysis:
1. The effects due to the non-idealness parameter and the adiabatic index of the gas, generally, do not change the trends of variations of the shock velocity and flow variables behind the cylindrical and spherical shock waves but they modify the numerical values of the shock velocity and flow variables from their values for the ideal gas case.
2. The shock velocity, pressure, particle velocity, speed of sound and change in entropy across the front increase whereas the adiabatic compressibility of medium decreases as the shock wave focuses at the centre of convergence. The density of gas behind the shock front remains independent with the propagation distance.
3. The shock velocity, pressure, density, particle velocity, speed of sound and change in entropy increase whereas the adiabatic compressibility decreases with increase in the value of non-idealness parameter of the gas.
4. The shock velocity, pressure, density, particle velocity, speed of sound and change in entropy decrease whereas the adiabatic compressibility increases with an increase in the value of adiabatic index of the gas.
5. The trends of variations of the shock velocity and the flow quantities just behind the shock wave in the real gases are similar to that of behind the shock wave in a perfect gas.

Except being of interest from a physicist's point of view, present and potential applications of the model developed in this paper are found in material science, nuclear science, astrophysics and space science. Strong imploding shock waves are of interest in material synthesis, where the phase, hardness or other characteristics of a material can be changed through shock wave compression e.g. synthesis of diamond from carbon. The model may be used to describe some of the overall features of the imploding shock waves in stars, stellar medium, double-detonation supernovae, astrophysical fluids, fusion reactions e.g. gamma-rays have been detected escaping from shock waves converging in deuterium, etc. This model is valid when the shock wave is away from the centre of convergence because at the centre the pressure becomes infinite which is not physical. The model's advantages lie in the fact that it is capable of describing the flow field just behind the strong imploding shocks in ideal as well as actual environments. The problem considered in the paper is of astrophysical interest, however, the methodology and analysis presented in the paper may be useful in many other physical systems which involve nonlinear hyperbolic wave propagation.

**Acknowledgement:** Thanks to my family for their support and encouragement during preparation of the paper.